\documentclass[journal]{IEEEtran}
\ifCLASSINFOpdf
\else
\fi

\usepackage{color}%
\usepackage{amsmath}%
\usepackage{graphicx} %
\begin{document}
%
% paper title
% Titles are generally capitalized except for words such as a, an, and, as,
% at, but, by, for, in, nor, of, on, or, the, to and up, which are usually
% not capitalized unless they are the first or last word of the title.
% Linebreaks \\ can be used within to get better formatting as desired.
% Do not put math or special symbols in the title.
\title{Latent Correlation Representation Learning for Brain Tumor Segmentation with Missing MRI Modalities}
%
%
% author names and IEEE memberships
% note positions of commas and nonbreaking spaces ( ~ ) LaTeX will not break
% a structure at a ~ so this keeps an author's name from being broken across
% two lines.
% use \thanks{} to gain access to the first footnote area
% a separate \thanks must be used for each paragraph as LaTeX2e's \thanks
% was not built to handle multiple paragraphs
%

\author{Tongxue Zhou,
        St\'{e}phane Canu,
        Pierre Vera,
        and Su Ruan*% <-this % stops a space
\thanks{* Corresponding author.}        
\thanks{T. Zhou is with LITIS, Apprentissage, INSA Rouen, 76800 Rouen, France, and also with the LITIS, QuantIF, Université de Rouen Normandie, 76183, Rouen, France (e-mail:
tongxue.zhou@insa-rouen.fr).}
\thanks{S. Canu is with LITIS, Normandie Univ, INSA Rouen, UNIROUEN, UNIHAVRE, France (e-mail: stephane.canu@insa-rouen.fr).}%
\thanks{P. Vera is with the Department of Nuclear Medicine, Henri Becquerel Cancer Center, 76038 Rouen, France, and also with the LITIS, QuantIF, Université de Rouen Normandie,
76183 Rouen, France (e-mail: pierre.vera@chb.unicancer.fr).}
\thanks{S. Ruan* is with the LITIS, QuantIF, Université de Rouen Normandie, 76183 Rouen, France (e-mail: su.ruan@univ-rouen.fr).}
\thanks{© 2021 IEEE. Personal use of this material is permitted. Permission from IEEE must be obtained for all
other uses, in any current or future media, including reprinting/republishing this material for advertising
or promotional purposes, creating new collective works, for resale or redistribution to servers or lists, or
reuse of any copyrighted component of this work in other works.}}

% note the % following the last \IEEEmembership and also \thanks - 
% these prevent an unwanted space from occurring between the last author name
% and the end of the author line. i.e., if you had this:
% 
% \author{....lastname \thanks{...} \thanks{...} }
%                     ^------------^------------^----Do not want these spaces!
%
% a space would be appended to the last name and could cause every name on that
% line to be shifted left slightly. This is one of those "LaTeX things". For
% instance, "\textbf{A} \textbf{B}" will typeset as "A B" not "AB". To get
% "AB" then you have to do: "\textbf{A}\textbf{B}"
% \thanks is no different in this regard, so shield the last } of each \thanks
% that ends a line with a % and do not let a space in before the next \thanks.
% Spaces after \IEEEmembership other than the last one are OK (and needed) as
% you are supposed to have spaces between the names. For what it is worth,
% this is a minor point as most people would not even notice if the said evil
% space somehow managed to creep in.

% The paper headers
\markboth{\small{\textbf{Accepted by IEEE Transactions on Image Processing (8 April 2021). Doi: 10.1109/TIP.2021.3070752.}}}%
{Shell \MakeLowercase{\textit{et al.}}: Latent Correlation Representation Learning for Brain Tumor Segmentation with Missing MRI Modalities}
% The only time the second header will appear is for the odd numbered pages
% after the title page when using the twoside option.
% 
% *** Note that you probably will NOT want to include the author's ***
% *** name in the headers of peer review papers.                   ***
% You can use \ifCLASSOPTIONpeerreview for conditional compilation here if
% you desire.

% If you want to put a publisher's ID mark on the page you can do it like
% this:
%\IEEEpubid{0000--0000/00\$00.00~\copyright~2015 IEEE}
% Remember, if you use this you must call \IEEEpubidadjcol in the second
% column for its text to clear the IEEEpubid mark.

% use for special paper notices
%\IEEEspecialpapernotice{(Invited Paper)}

% make the title area
\maketitle
% As a general rule, do not put math, special symbols or citations
% in the abstract or keywords.
\begin{abstract}
Magnetic Resonance Imaging (MRI) is a widely used imaging technique to assess brain tumor. Accurately segmenting brain tumor from MR images is the key to clinical diagnostics and treatment planning. In addition, multi-modal MR images can provide complementary information for accurate brain tumor segmentation. However, it's common to miss some imaging modalities in clinical practice. In this paper, we present a novel brain tumor segmentation algorithm with missing modalities. Since it exists a strong correlation between multi-modalities, a correlation model is proposed to specially represent the latent multi-source correlation. Thanks to the obtained correlation representation, the segmentation becomes more robust in the case of missing modality. First, the individual representation produced by each encoder is used to estimate the modality independent parameter. Then, the correlation model transforms all the individual representations to the latent multi-source correlation representations. Finally, the correlation representations across modalities are fused via attention mechanism into a shared representation to emphasize the most important features for segmentation. We evaluate our model on BraTS 2018 and BraTS 2019 dataset, it outperforms the current state-of-the-art methods and produces robust results when one or more modalities are missing.
\end{abstract}

% Note that keywords are not normally used for peerreview papers.
\begin{IEEEkeywords}
Brain tumor segmentation, multi-modal, missing modalities, fusion, latent correlation representation, deep learning.
\end{IEEEkeywords}

% For peer review papers, you can put extra information on the cover
% page as needed:
% \ifCLASSOPTIONpeerreview
% \begin{center} \bfseries EDICS Category: 3-BBND \end{center}
% \fi
%
% For peerreview papers, this IEEEtran command inserts a page break and
% creates the second title. It will be ignored for other modes.
\IEEEpeerreviewmaketitle

\section{Introduction}
\begin{figure}[htb]
\centering
\centerline{\includegraphics[width=8cm]{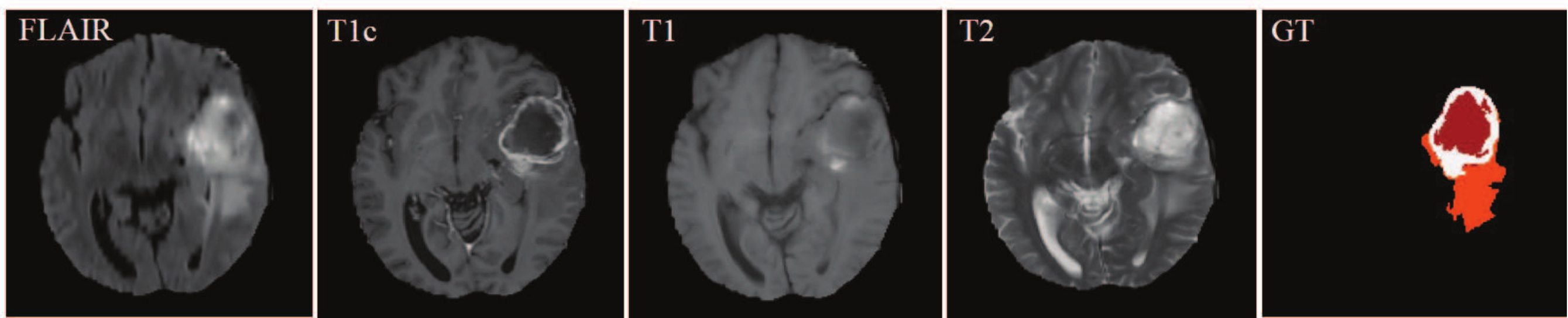}}
\caption{Example of data from a training subject. The first four images from left to right show the MRI modalities: Fluid Attenuation Inversion Recovery (FLAIR), contrast enhanced T1-weighted (T1c), T1-weighted (T1), T2-weighted (T2) images, and the fifth image is the ground truth labels, Net\&Ncr is shown in red, edema in orange and enhancing tumor in white, Net refers non-enhancing tumor and Ncr necrotic tumor.}
\label{fig1}
\end{figure}

% The very first letter is a 2 line initial drop letter followed
% by the rest of the first word in caps.
% 
% form to use if the first word consists of a single letter:
% \IEEEPARstart{A}{demo} file is ....
% 
% form to use if you need the single drop letter followed by
% normal text (unknown if ever used by the IEEE):
% \IEEEPARstart{A}{}demo file is ....
% 
% Some journals put the first two words in caps:
% \IEEEPARstart{T}{his demo} file is ....
% 
% Here we have the typical use of a "T" for an initial drop letter
% and "HIS" in caps to complete the first word.
\IEEEPARstart{B}{rain} tumor is one of the most aggressive cancers in the world, early diagnosis of brain tumors plays an important role in clinical assessment and treatment planning of brain tumors. Magnetic Resonance Imaging (MRI) is a widely used imaging technique to assess brain tumor, because it offers a good soft tissue contrast without radiation. The commonly used sequences are T1-weighted, contrast enhanced T1-weighted (T1c), T2-weighted and Fluid Attenuation Inversion Recovery (FLAIR) images. Different sequences can provide complementary information to analyze different subregions of gliomas. For example, T2 and FLAIR highlight the tumor with peritumoral edema, designated whole tumor. T1 and T1c highlight the tumor without peritumoral edema, designated tumor core. An enhancing region of the tumor core with hyper-intensity can also be observed in T1c, designated enhancing tumor core \cite{zhou2019review}. In this work, we refer to  these images of different sequences as  modalities. Therefore applying multi-modal images can reduce the information uncertainty and improve clinical diagnosis and segmentation accuracy. The four MRI modalities and the related ground truth are shown in Fig. \ref{fig1}, we can see FLAIR can give important information about whole tumor, while T1c can provide more information about tumor core (including enhancing tumor and Net\&Ncr regions).

Traditional brain tumor segmentation methods such as probability theory \cite{lapuyade2017segmenting}, kernel feature selection \cite{zhang2011kernel}, belief function \cite{lian2018joint}, random forests \cite{zikic2012decision}, conditional random fields \cite{yu2018semi} and support vector machines \cite{bauer2011fully} have achieved a great success in recent years. However, brain tumor segmentation is still a challenging task, especially in the case of missing some modalities. There are three main reasons: (1) The different brain anatomy structures between patients. (2) The size, shape, and texture of gliomas varies from patients to patients. (3) The variability in intensity range and low contrast in qualitative MR imaging modalities. This is particularly true for brain tumor segmentation, where the tumor contour is fuzzy due to low contrast.

Segmentation of brain tumor by experts is expensive and time-consuming. Recently, with a strong feature learning ability, deep learning-based approaches have become more prominent for brain tumor segmentation. Zhou et al. \cite{zhou2020one} presented One-pass Multi-task Network (OM-Net) for brain tumor segmentation which is tailored to handle the class imbalance problem. Cui et al. \cite{cui2018automatic} proposed a cascaded deep learning convolutional neural network consisting of two sub-networks. The first network is to define the tumor region from an MRI slice and the second network is used to label the defined tumor region into multiple sub-regions. Mlynarski et al. \cite{mlynarski20193d}
introduced a CNN-based model to efficiently combine the advantages of the short-range 3D context and the long-range 2D context for brain tumor segmentation. Wang et al. \cite{wang2019multimodal} proposed a novel 2D fully convolution segmentation network WRN-PPNet based on the pyramid pooling module. Kamnitsas et al. \cite{kamnitsas2017efficient} proposed an efficient multi-scale 3D CNN with fully connected CRF for brain tumor segmentation. Havaei et al. \cite{havaei2017brain} implemented a two-pathway architecture that learns about the local details of the brain tumor as well as the larger context feature. Wang et al. \cite{wang2017automatic} proposed to use three networks to hierarchically and sequentially segment substructures of brain tumor. Kamnitsas et al. \cite{kamnitsas2017ensembles} proposed the Ensembles of Multiple Models and Architectures (EMMA) for robust performance, which won the first position in BraTS 2017 competition. Considering the limited medical image data but rich information in modality property, Zhang et al. \cite{zhang2020cross} proposed a novel cross-modality deep feature learning framework for multi-modal brain tumor segmentation. It consists of a cross-modality feature transition process and a cross-modality feature fusion process. The former one is to learn rich feature representations by transiting knowledge across different modality data, the latter one is to fuse knowledge from different modality data. Chen et al. \cite{chen2019aggregating} proposed a network which aggregates multi-scale predictions from the decoder part of 3D U-Net for brain tumor segmentation. Hamghalam et al. \cite{hamghalam2019brain} introduced a synthetic segmentation framework by translating FLAIR MR images into high-contrast synthetic MR ones for segmentation, which can decrease the number of real channels in multi-modal brain tumor segmentation. Cheng et al. \cite{cheng2019memory} proposed a novel memory-efficient cascade 3D U-Net for brain tumor segmentation, which integrated the multi-scale information fusion mechanism, cascade strategy as well as the combination of edge loss and weighted dice loss.

The above approaches require the complete set of the modalities. However, the imaging modalities are often incomplete or missing in clinical practice. In this paper, we focus on brain tumor segmentation algorithm with missing modalities. Currently, there are a number of methods proposed to deal with the missing modalities in medical image segmentation, which can be broadly grouped into three categories: (1) training a model on all possible subsets of the modalities, which is complicated and time-consuming. (2) synthesizing missing modalities and then use the complete imaging modalities to do the segmentation, while it requires an additional network for synthesis and the quality of the synthesis can directly affect the segmentation performance. (3) fusing the available modalities in a latent space to learn a shared feature representation, then project it to the segmentation space. This approach is more efficient than the first two methods, because it doesn't need to learn a number of possible subsets of the multi-modalities and will not be affected by the quality of the synthesized modality. 

Recently, there are a lot of segmentation methods based on exploiting latent feature representation for missing modalities. The problem is what kind of latent features need to be learned and how to learn them. The current state-of-the-art network architecture is from Havaei, the proposed HeMIS \cite{havaei2016hemis} learns the feature representation of each modality separately, and then the first and second moments are computed across individual modalities for estimating the final segmentation. However, computing mean and variance over individual representations can't learn the shared latent representation. Lau et al. \cite{lau2019unified} introduced a unified representation network that maps a variable number of input modalities into a unified representation by using mean function for segmentation, while averaging the latent representations could loss partially important information. Chen et al. \cite{chen2019robust} used feature disentanglement to decompose the input modalities into content code and appearance code, and then the content code are fused via a gating strategy into a shared representation for segmentation. While the approach is more complex and time-consuming, because it requires two encoders for each modalities. And their proposed fusion method only re-weights the content code from spatial-wise without channel-wise. Shen et al. \cite{shen2019brain} used adversarial loss to form a domain adaptation model to adapt feature maps from missing modalities to the one from full modalities, which can only cope with the one-missing modality situation. In  this work, we present a correlation model to represent a latent correlated relationship  between features of each modality. Our idea is to use deep learning to discover this latent correlated relationship, helping the segmentation even if some modalities are missing.

For multi-modal medical image segmentation task, the fusion strategy takes an important role to achieve an accurate segmentation result, as presented in \cite{zhou2019review}. Since not all features extracted from the network are useful for segmentation. Therefore, it is necessary to find an effective way to fuse features, and find the most informative features for segmentation. To this end, we propose to use the attention mechanism, which can be viewed as a tool being capable to take into account the most informative feature representation. Channel attention modules and spatial attention modules are the commonly used attention mechanisms. The former one, channel attention modules, use attention mechanism to select meaningful features at each channel axis. For example, Hu et al. \cite{hu2018squeeze} introduced the Squeeze and Excitation (SE) block to perform dynamic channel-wise feature recalibration to improve the representational power of a network. Li et al. \cite{li2018pyramid} proposed to combine attention mechanism and spatial pyramid to extract precise dense features for pixel labeling in semantic segmentation. Oktay et al. \cite{oktay2018attention} proposed an attention U-net, which uses a channel attention mechanism to fuse the high-level and low-level features for CT abdominal segmentation. The latter one, spatial attention modules, calculate the feature representation in each position by weighted sum of the features of all other positions. For example, Roy et al. \cite{roy2018concurrent} proposed to use both spatial and channel SE blocks (scSE) and demonstrated that scSE blocks can yield an improvement on three different FCNN architectures. Currently, Roy et al. \cite{roy2020squeeze} incorporated scSE blocks to the few-shot segmentation task for the segmentation of volumetric medical images with only a few annotated slices. Fu et al. \cite{fu2019dual} presented a dual attention network using the channel and spatial attention mechanisms to adaptively integrate local semantic features with global dependencies for scene segmentation. However, the methods mentioned above evaluated the attention mechanism only on the single-modal image dataset and didn't consider the fusion issue on the multi-modal medical images. To this end, we applied the attention mechanism to the multi-modal segmentation network to achieve the effective feature fusion.

The challenge of segmentation on missing modalities is to learn a shared latent representation, which can take any subset of the image modalities and produce robust segmentation. To effectively learn the latent representation of individual representations, in this paper, we propose a novel brain tumor segmentation network to deal with the absence of imaging modalities. The main contributions of our method are three folds: 1) A correlation model is introduced to discover the latent multi-source correlation representation. 2) A fusion strategy based on attention mechanism with obtained correlation representation is proposed to segment brain tumor in the case of missing modalities.  3) To the best of our knowledge, this is the first multi modal segmentation network which is capable of describing the latent multi-source correlation representation and then using it to help segmentation for missing data. 

The paper is organized as follows: Section \ref{sec2} offers an overview of this work and details our model, Section \ref{sec3} describes experimental setup and used datasets, Section \ref{sec4} presents the experimental results and discusses relevant finding. Section \ref{sec5} discusses the proposed method and concludes this work.

\section{Method}
\label{sec2}
\begin{figure}[htb]
\includegraphics[width=9cm]{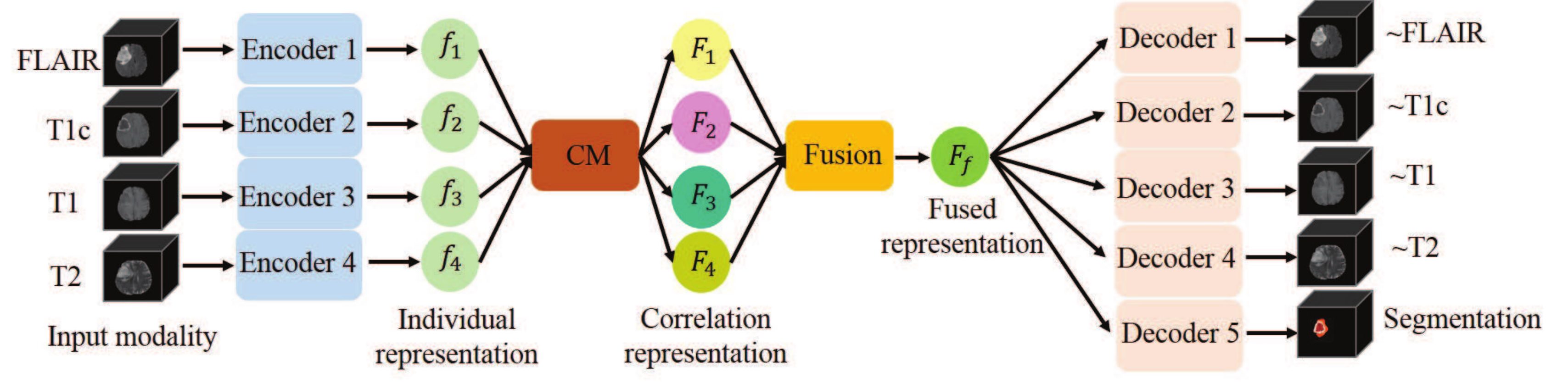}
\caption{A schematic overview of our network. Each input modality is encoded by individual encoder to obtain the individual representation. The proposed Correlation Model (CM) and Fusion block (Fusion) project the individual representations into a fused representation, which is finally decoded to form the reconstruction modalities and the segmentation result. Full details of each component are given in the main text.}
\label{fig2}
\end{figure}

Our network is based on the U-Net architecture \cite{ronneberger2015u}, in which we integrate our fusion strategy and correlation model. To be robust to the absence of modalities, we adapt it to multi-encoder based framework. It first takes 3D available modalities as inputs in each encoder. The independent encoders can not only learn modality-specific feature representation, but also can avoid the false-adaptation between modalities. To take into account the strong correlation between multi modalities, we propose a correlation model, named CM, to discover the correlation between modalities. Then the correlation representations across modalities are fused via attention mechanism, named Fusion, to emphasize the most discriminative representation for segmentation. Finally, the fused latent representation is decoded to form the final segmentation result. The network architecture scheme is depicted in Fig. \ref{fig2}.

\subsection{Modeling the latent multi-source correlation}
\begin{figure}[htb]
\centering
\includegraphics[width=8cm]{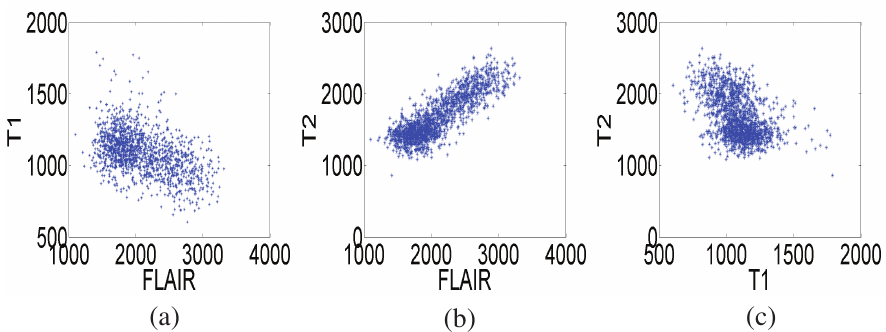}
\caption{Joint intensity distribution of MR images: (a) FLAIR-T1, (b) FLAIR-T2 and (c) T1-T2.}
\label{fig3}
\end{figure}

\begin{figure}[htb]
\centering
\includegraphics[width=9cm]{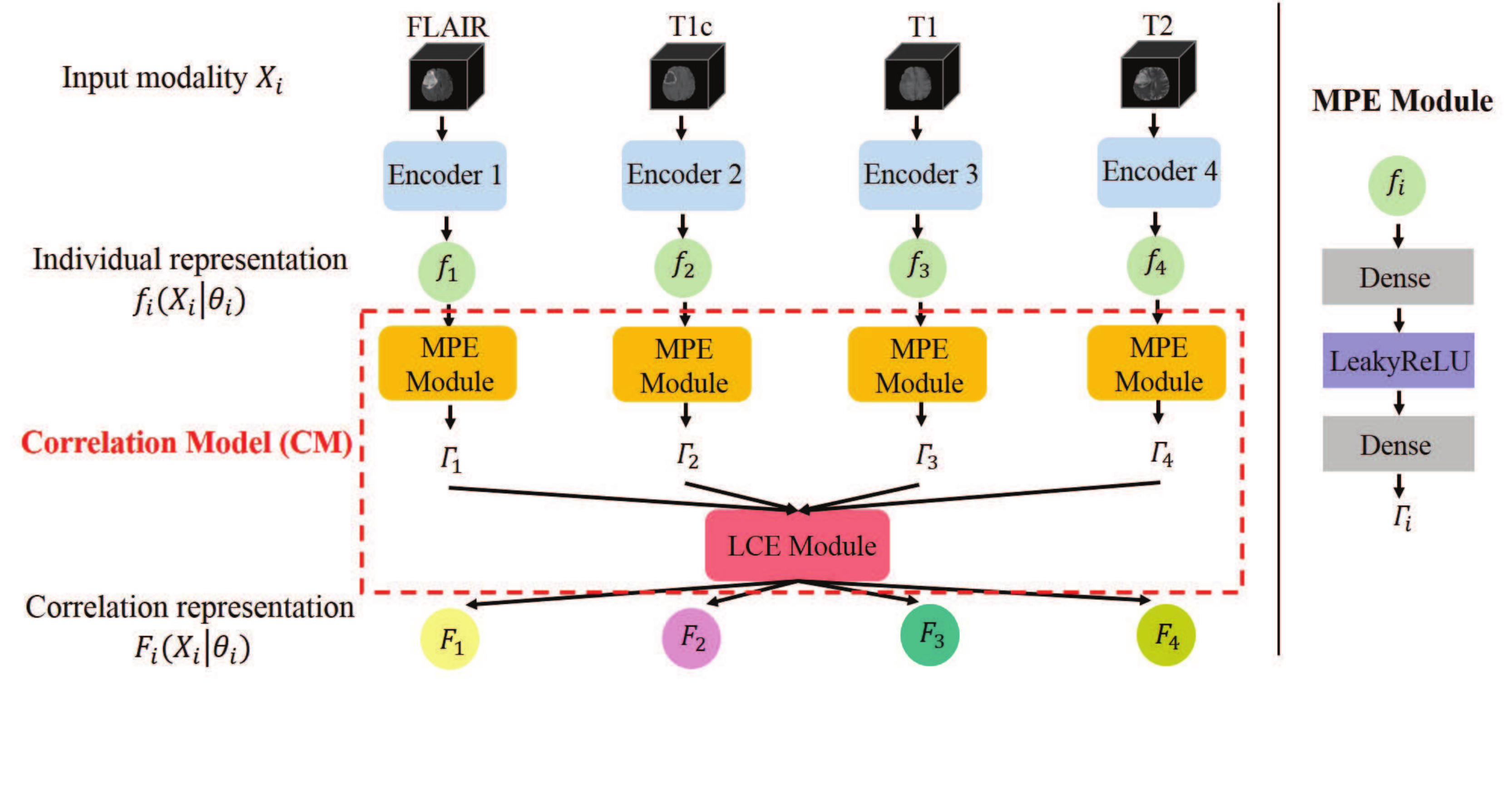}
\caption{Architecture of correlation model. MPE Module first maps the individual representation $f_i(X_i|\theta_i)$ to a set of independent parameters $\Gamma_i$, under these parameters, LCE Module transforms all the individual representations to form a  latent multi-source correlation representation $F_i(X_i|\theta_i)$.}
\label{fig4}
\end{figure}

Inspired by a fact that, there is strong correlation between multi MR modalities, since the same scene (the same patient) is observed by different modalities \cite{lapuyade2017segmenting}. From Fig. \ref{fig3} presenting joint intensity distribution of each pair of MR images corresponding respectively to FLAIR-T1, FLAIR-T2 and T1-T2, we can observe a strong correlation in intensity distribution between each two modalities. To this end, it's reasonable to assume that a strong correlation also exists in latent representation between modalities. And we introduce a Correlation Model (CM) (see Fig. \ref{fig4}) to discover the latent correlation between the four modalities: FLAIR, T1, T1c and T2. The CM consists of two modules: Model Parameter Estimation Module (MPE Module) and Linear Correlation Expression Module (LCE Module). Each input modality $\{X_i\}$, where $i=\{1,2,3,4\}$, is first input to the independent encoder to learn the modality-specific representation $f_i(X_i|\theta_i)$, where $\theta_i$ denotes the parameters used in $i$th encoder, such as the number of filters, the kernel size of filter and the rate of dropout, which aid the encoder to obtain the most discriminative representation. Then, MPE Module, a network with two fully connected layers and LeakyReLU, maps the modality-specific representation $f_i(X_i|\theta_i)$ to a set of independent parameters $\Gamma_i =\{\alpha_i, \beta_i, \gamma_i, \delta_i\}$, which is unique for each modality. Finally, the correlation representation $F_i(X_i|\theta_i)$ can be obtained via LCE Module (Equation \eqref{eq1}). It is noted that the number of latent multi-source correlation representation equals to the number of modalities, and the differences among the correlation representations are the weights in each correlation representation, which are related to the individual modalities.

\begin{equation}
\begin{split}
    &F_i(X_i|\theta_i) = \alpha_i \odot f_j(X_{j}|\theta_{j})+\beta_i \odot f_k(X_{k}|\theta_{k})+\\
&\gamma_i \odot f_m(X_{m}|\theta_{m})+\delta_i,
    (i \neq j \neq k \neq m)
\label{eq1}
\end{split}
\end{equation}

Utilizing the correlation model, we can use the available modalities to describe the missing one, which allows the segmentation network to have strong capability to deal with missing data.

\subsection{Fusion}
\label{sec2.4}
The purpose of fusion is to emphasize the most important features from different modalities to highlight regions that are greatly relevant to brain tumor. One simple way to obtain the fused representation is to average over them. 
\begin{equation}
    F_f= mean(F_1, ..., F_n)
\end{equation}
where $F_n$ denotes the $n$th individual representation, $F_f$ denotes the fused representation. 

While the average operation makes each representation contribute equally to the fused representation, which could lose some valuable information in the latent representation. 

Another way to combine the individual representations into a fused representation is to maximum the most important representation from all the representations. 
\begin{equation}
    F_f= max(F_1, ..., F_n)
\end{equation}

The maximum fusion method only takes into account the most discriminative representation, while it could ignore the contributions of other representations. 

\begin{figure}[htb]
\centering
\includegraphics[width=9cm]{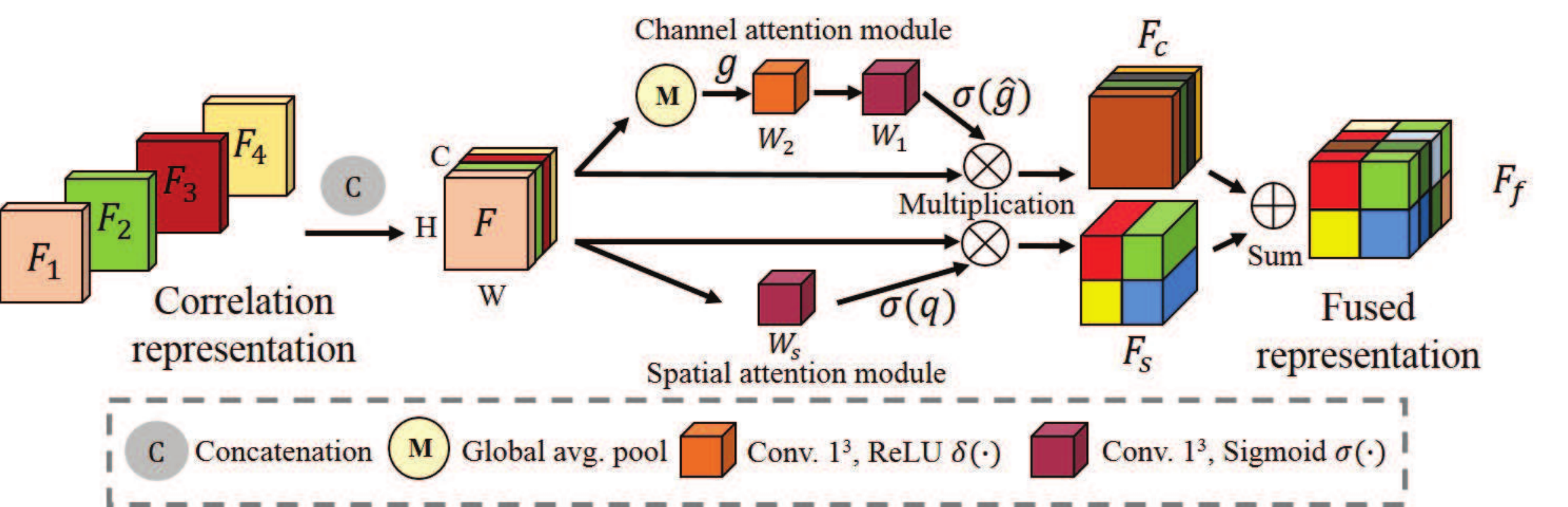}
\caption{The architecture of fusion block. The correlation representations ($F1$, $F2$, $F3$, $F4$) are first concatenated as the input of the attention mechanism $F$. Then, they are recalibrated along channel attention and spatial attention modules to achieve the $F_c$ and $F_s$. Finally, they are added to obtain the fused representation $F_f$.}
\label{fig5}
\end{figure}

In multi-modal segmentation, not only the different modalities have various contributions but also different spatial location in each modality can give different weights on feature representation for segmentation. To this end, we introduced attention mechanism to the fusion block, described in Fig. \ref{fig5}. The correlation representations obtained from CM are first concatenated as the input representation $F= [F_1, F_2, F_3, F_4]$, $F_k\in R^{H\times W}$. Note that, in the lowest level of the network, there are four correlation representations for the fusion block, in the other levels, the result of the previous level is also concatenated with the modality-specific representation to obtain five input representations, in the following, we just describe the fusion block in the lowest level, because the other levels are similar.

In the channel attention module, a global average pooling is first performed to produce a tensor $g\in R^{1\times1\times 4}$, which represents the global spatial information of the representation, with its $k^{th}$ element
\begin{equation}
    g_k= \frac{1}{H\times W}\sum_i^H\sum_j^W F_k(i,j)
\end{equation}

Then, two fully-connected layers are applied to encode the channel attention dependencies, $\hat{g}= W_1(\delta(W_2 g))$, with $W_1\in R^{4\times2}$, $W_2\in R^{2\times4}$, being weights of two fully-connected layers and the ReLU operator $\delta(\cdot)$. $\hat{g}$ is then passed through the sigmoid layer to obtain the channel attention weights, which will be applied to the input representation $F$ through multiplication to achieve the channel attention representation $F_c$, the $\sigma(\hat{g_k})$ indicates the importance of the $i$ channel of the representation: \begin{equation}
    F_c= [\sigma (\hat{g_1})F_1, \sigma(\hat{g_2})F_2, \sigma(\hat{g_3})F_3, \sigma(\hat{g_4})F_4,]
\end{equation}

In the spatial attention module, the representation can be considered as $F= [F^{1,1}, F^{1,2}, ... , F^{i,j},..., F^{H,W}]$, $F^{i,j}\in R^{1\times1\times4}$, $i\in{1, 2,..., H}$, $j\in{1, 2,...,W}$, and then a convolution operation $q=W_s\star F$, $q\in R^{H\times W}$  with weight $W_s\in R^{1\times1\times4\times1}$, is used to squeeze the spatial domain, and to produce a projection tensor, which represents the linearly combined representation for all channels for a spatial location. The tensor is finally passed through a sigmoid layer to obtain the spatial attention weights and to achieve the spatial attention representation $F_s$, the $\sigma(q_{i,j})$ that indicates the importance of the spatial information $(i, j)$ of the representation:
\begin{equation}
    F_s=[\sigma(q_{1,1})F^{1,1},  ... , \sigma(q_{i,j})F^{i,j}, ... , \sigma(q_{H,W})F^{H,W}]
\end{equation}

The fused representation representation is obtained by adding the channel attention representation and spatial attention representation:
\begin{equation}
    F_f=F_c+F_s 
\end{equation}

The proposed fusion method can be directly adapted to any multi modal fusion problem, and it encourages the network to learn more meaningful representation along spatial attention and channel attention, which is superior than simple mean and maximum fusion methods.

\subsection{Network Architecture and Learning Process}
\begin{figure*}[htb]
\centering
\includegraphics[width=18cm]{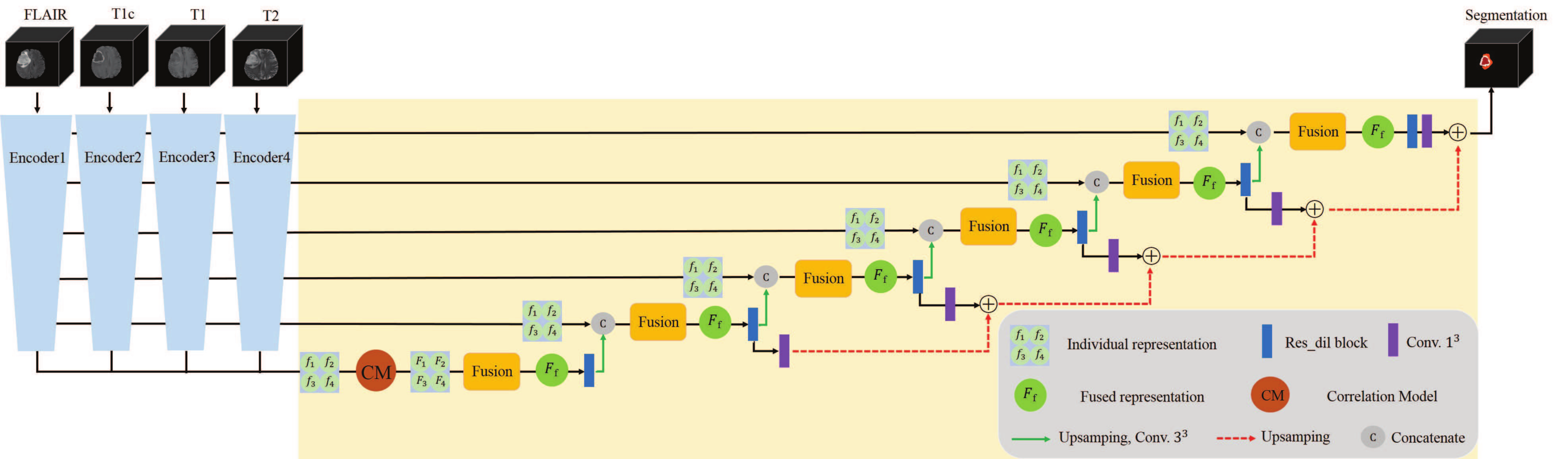}
\caption{Proposed segmentation network framework. Each imaging modality (FLAIR, T1, T1c, T2) is first encoded by individual encoder to obtain the modality-specific representation ($f_1$, $f_2$, $f_3$, $f_4$). Then these individual representations are transformed by CM to form the dependent representation ($F_1$, $F_2$, $F_3$, $F_4$), which are integrated by the following fusion block. Finally, the fused latent representation $F_f$ is decoded to form the final segmentation. Here only four encoders and the target segmentation decoder are shown.}
\label{fig6}
\end{figure*}

The detailed network architecture framework is illustrated in Fig. \ref{fig6}. The encoder is used to obtain the modality-specific representation, which includes convolutional block, res\_dil block followed by skip connection. In order to maintain the spatial information, we use convolution with stride=2 to replace pooling operation. It's likely to require different receptive field when segmenting different regions in an image. And standard U-Net can’t get enough semantic features due to the limited receptive field. Inspired by dilated convolution, we use residual block with dilated convolutions (rate = 2, 4) (res\_dil block) on both encoder part and decoder part to obtain features at multiple scale. It can obtain more extensive local information to help retain information and fill details during training process. All convolutions are $3\times3\times3$ and the number of filter is increased from 8 to 256. Each decoder level begins with up-sampling layer followed by a convolution to reduce the number of features by a factor of 2. Then the upsampled features are combined with the features from the corresponding level of the encoder part using concatenation. After the concatenation, we use the res\_dil block to increase the receptive field. In addition, we employ deep supervision \cite{isensee2017brain} for the segmentation decoder by integrating segmentation layers from different levels to form the final network output. 

\subsection{The choices of loss function}
The network is trained by the overall loss function as follow:
\begin{equation}
    \ L_{total}= L_{dice} + \lambda L_{rec}
\end{equation}
\noindent where $\lambda$ is the trade-off parameters weighting the importance of each component, which is set as 1 in our experiment. 

For segmentation, we use dice loss to evaluate the overlap rate of prediction results and ground truth.
\begin{equation}
    \ L_{dice}=1-2\frac{\sum_{i=1}^N\sum_{j=1}^C p_{ij} g_{ij}+\epsilon} {\sum_{i=1}^N\sum_{j=1}^C p_{ij} + g_{ij}+\epsilon}
\end{equation}
\noindent where $N$ is the set of all examples, $C$ is the set of the classes, $p_{ij}$ is the predicted probability that pixel $i$ is of the tumor class $j$. $g_{ij}$ is the real probability that pixel $i$ is of the tumor class $j$, and $\epsilon$ is a small constant to avoid dividing by 0.
%and $p_{i\overline{j}}$ is the probability that pixel $i$ is of the non-tumor class $j$. and $g_{i\overline{j}}$
For reconstruction, we use Mean Absolute Error (MAE) to match each reconstruction image to the input image. 
\begin{equation}
 L_{rec} = \sum_{i=1}^{n} MAE{\lVert R_i(z)- x_i \lVert}_1
\end{equation}
\noindent where $n$ denotes the number of modality, in this task, $n=4$, $R$ denotes reconstruction path, $z$ is the fused representation obtained by fusion block in the last layer, $x_i$ is the input modality.

\section{Experimental setup}
\label{sec3}
\subsection{Data and preprocessing}
The datasets used in the experiments come from BraTS 2018 and BraTS 2019 dataset \cite{bakas2018identifying}. The BraTS 2018 dataset has a training set which includes 285 cases, and a validation set including 66 cases with hidden ground-truth. The BraTS 2019 dataset has a training set which includes 335 cases, and a validation set including 125 cases with hidden ground-truth. In both datasets, each case has four image modalities including T1, T1c, T2 and FLAIR. Following the challenge, four intra-tumor structures (edema, enhancing tumor, necrotic and non-enhancing tumor core) have been grouped into three mutually inclusive tumor regions: 
\noindent (a) The whole tumor region (WT), consisting of all tumor tissues. 
\noindent (b) The tumor core region (TC), consisting of the enhancing tumor, necrotic and non-enhancing tumor core. 
\noindent (c) The enhancing tumor region (ET). 

The provided data have been pre-processed by organisers: co-registered to the same anatomical template, interpolated to the same resolution ($1 mm^3$) and skull-stripped. The ground truth have been manually labeled by experts. We did additional pre-processing with a standard procedure. To exploit the spatial contextual information of the image, we use 3D image, crop and resize it from $155 \times 240 \times240$ to $128 \times 128 \times128$. The N4ITK \cite{avants2009advanced} method is used to correct the distortion of MRI data, and intensity normalization is applied to normalize each modality of each patient. 

\subsection{Implementation details}
Our network is implemented in Keras with a single Nvidia GPU Quadro P5000 (16G). The model is optimized using the Adam optimizer (initial learning rate = 5e-4) with a decreasing learning rate factor 0.5 with patience of 10 epochs, and early stopping is employed to avoid over-fitting if the validation loss isn’t improved over 50 epochs. We randomly split the dataset into 80\% training and 20\% testing. All the results are obtained by online evaluation platform.

\subsection{Evaluation metrics}
To evaluate the proposed method, we employ three widely used metrics: Dice Score, Sensitivity and Hausdorff Distance to obtain quantitative measurements of the segmentation accuracy.

\noindent 1) Dice Score: It is designed to evaluate the overlap rate of prediction results and ground truth. Dice ranges from 0 to 1, and the better predict result will have a larger Dice value.

\begin{equation}
 Dice\ Score= \frac {2TP}{2TP+FP+FN}
\end{equation}

\noindent 2) Sensitivity (also called the true positive rate, the recall): It measures the proportion of actual positives that are correctly identified:

\begin{equation}
Sensitivity=\frac{TP}{TP+FN}
\end{equation}

\noindent where $TP$ represents the number of true positive voxels, $FP$ the number of false positive voxels, and $FN$ the number of false negative voxels. 

\noindent 3) Hausdorff Distance (HD): It is computed between boundaries of the prediction results and ground-truth, it is an indicator of the largest segmentation error. The better predict result will have a smaller HD value.

\begin{equation}
  HD=\max\{sup_{r_\in{\partial R}}d_m(s,r),sup_{s_\in\partial S}d_m(r,s)\}
\end{equation}

\noindent where $\partial S$ and $\partial R$ are the sets of tumor border voxels for the predicted and the real annotations, and $d_m(v,v)$ is the minimum of the Euclidean distances between a voxel $v$ and voxels in a set $v$.

\section{Experiment results}
\label{sec4}
Here, we conduct a series of comparative experiments to demonstrate the effectiveness of our proposed method and compare it to other approaches. In Section \ref{4.1.1}, we first perform experiments to see the importance of our proposed components and demonstrate that adding the proposed components can enhance the segmentation performance. In Section \ref{4.1.2}, we compare our method with the state-of-the-art methods on full modalities. In Section \ref{4.1.3}, we demonstrate that the proposed CM can discover the latent representation between modalities to make the model robust on missing data. And it gives a better performance than the state-of-art approach HeMIS. In Section \ref{4.2}, the qualitative experiment results further demonstrate that our proposed method can achieve a promising segmentation result on complete and incomplete modalities. In Section \ref{4.3}, we visualized the feature maps of the attention based fusion block to demonstrate that it can encourage the network to learn meaningful representations for segmentation.

\subsection{Quantitative analysis}
\label{ssec:subsubhead}
To prove the effectiveness of our network architecture, we first carry out the ablation experiments to evaluate our method and demonstrate the effectiveness of the proposed components on fully modalities, and then we compare our model with the state-of-the-art methods on fully modalities. Finally, we analyze the robustness of our method in the case of missing modalities.

\subsubsection{Evaluation of our method on full modalities}
\label{4.1.1}
To show the performance of our method, and see the importance of the proposed components in our network, including fusion block, reconstruction decoder and CM, we did the ablation studies on both BraTS 2018 and 2019 datasets, and the results are shown in Table. \ref{tab1} and Table. \ref{tab2}, respectively. We denote the original network without fusion block, CM and reconstruction decoders as baseline. From Table. \ref{tab1}, we can observe that the baseline method achieves average Dice Score, average Hausdorff Distance and average Sensitivity of 76.7, 8.9, 76.9, respectively. From Table. \ref{tab2}, we can see that the baseline method achieves average Dice Score, average Hausdorff Distance and average Sensitivity of 77.4, 9.6, 71.7, respectively. When the fusion block is applied to the network, we can see an improvement of average Dice Score, average Hausdorff Distance and also the average Sensitivity. The major reason is that the fusion block can help to emphasize the most important representations from all the different modalities to boost the segmentation result. In addition, when the reconstruction decoders are integrated to the network, the average Dice Score and average Sensitivity are further improved. Another advantage of our method is using the correlation representation, which can discover the latent correlation representation between modalities to make the model achieve better segmentation. From Table. \ref{tab1}, we can observe that the proposed CM can improve the baseline with 2.6\% in the terms of average Dice Score, 20.2\% in the terms of average Hausdorff Distance and 3.0\% in the terms of average Sensitivity on BraTS 2018 dataset. From Table. \ref{tab2}, we can observe the similar improvements of the proposed CM on BraTS 2019 dataset, which enhances the baseline with 2.5\% in the terms of average Dice Score, 27.1\% in the terms of average Hausdorff Distance and 9.6\% in the terms of average Sensitivity.

\subsubsection{Comparison with the state-of-the-art methods on full modalities}
\label{4.1.2}
We compare our proposed method with the state-of-the-art methods on BraTS 2018 and BraTS 2019 online validation sets, respectively. We first predict the segmentation results on local machine and then submitted them on the online evaluation platform to obtain the evaluation results. The results are presented in Table. \ref{tab3} and Table. \ref{tab4}, to make the comparison justify, we exclude the methods using post-processing.

The compared methods on BraTS 2018 datasets are introduced as following:

(1) Ronneberger et al. \cite{ronneberger2015u} proposed the U-Net, a widely used and effective approach for medical image segmentation. 

(2) Tuan et al. \cite{tuan2018brain} proposed to combine the Bit-plane method \cite{gonzalez2004digital} and U-Net architecture for brain tumor segmentation.

(3) Hu et al. \cite{hu2018brain} proposed a Multi-level Up-sampling Network (MU-Net) for brain tumor segmentation, which learns the feature representations from transverse, sagittal and coronal views and then fuses them to obtain the segmentation result.

(4) Myronenko et al. \cite{myronenko20183d} proposed to use the additional variational autoencoder (VAE) branch in U-Net architecture to regularize the shared encoder to cope with the limited data, and also to improve the segmentation performance.

From the compared results presented in Table. III, we can observe that the plain U-Net has an unsatisfied performance on all the tumor regions. The high HD values illustrate that the method has the large segmentation errors on all the tumor regions, while other improved U-Net based methods have much better segmentation results. Secondly, we can observe that our proposed method achieves the second best Dice Score across all the tumor regions, the second best Hausdorff Distance on tumor core, and the best average Sensitivity, which demonstrates the effectiveness of our proposed method. Thirdly, it can be seen that the best method is from \cite{myronenko20183d}, which achieves 90.4, 85.9 and 81.4 in terms of Dice Score on whole tumor, tumor core and enhancing tumor regions, respectively. However, it can only segment the brain tumors on full modalities, while our method is trained to do the segmentation when modalities are missing. In addition, the method \cite{myronenko20183d} fed a very large patch size (160 voxels $\times$ 192 voxels $\times$ 128 voxels) into the network and used 32 initial convolution filters. In addition, it required a 32GB GPU to train the model, which is computationally expensive. In contrast, our method used only 8 initial filters and exhibited a reasonable performance. A 16GB GPU is sufficient to conduct our experiments.

On BraTS 2019 dataset, we added 4 compared methods which are from:

(1) Agravat et al. \cite{agravat2019brain}  proposed an encoder-decoder based architecture along with dense connection for brain tumor segmentation.

(2) Boutry et al. \cite{boutry2019using} proposed to use separated inputs for multi-modal brain tumor segmentation, which forces the network to optimize different weights for each modality during the learning procedure.

(3) Kim et al. \cite{kim2019two} presented a two-step convolutional neural network for brain tumor segmentation. First, an initial segmentation probability map is obtained using the ensemble 2D U-Nets. Secondly, a 3D U-Net takes both the MR image and initial segmentation map as inputs to generate the final segmentation.

(4) Amian et.al \cite{amian2019multi} introduced to use two parallel networks with different resolutions for brain tumor segmentation.

Inspired by \cite{myronenko20183d}, the best method in BraTS 2019 challenge is presented by Jiang et al. \cite{jiang2019two} who proposed a novel two-stage cascaded U-Net to segment the substructures of brain tumors from coarse to fine. In the first stage, a variant of U-Net is used to get a coarse prediction. In the second stage, the coarse segmentation map is fed together with the raw images into the second U-Net to obtain a more accurate segmentation map. Finally, the well-designed post-processing is applied to refine the segmentation result. This work won the first place in the BraTS 2019 challenge segmentation task. It achieves 85.7 in the terms of average Dice Score, which is chosen from an ensemble of 12 models training from scratch. However, we don’t put it in the Table. \ref{tab4}, since this method used a post-processing, while other compared methods didn’t do it.

From the results in Table. \ref{tab4}, we can observe that our proposed method outperforms \cite{agravat2019brain, boutry2019using, kim2019two} in the terms of the average Dice Score, and ours obtain the best Hausdorff Distance across all the tumor regions, and also the second best Sensitivity on whole tumor and enhancing tumor regions.

\subsubsection{Comparison with the state-of-the-art methods on missing modalities}
\label{4.1.3}
The main advantage of our method is using the correlation representation, which can discover the latent correlation representation between modalities to make the model robust in the case of missing modalities. To demonstrate the effectiveness of the proposed CM and evaluate our method’s robustness performance on missing modalities, we compare it with WoCM, a specific case of our method without CM, and the state-of-the-art method \cite{havaei2016hemis} on BraTS 2018 and 2019 datasets. Since the authors of the method \cite{havaei2016hemis} didn't provide the available code, therefore, we can't compare the results on BraTS 2019 dataset. It is noted that the reported results of \cite{havaei2016hemis} on BraTS dataset 2018 are taken from the work in \cite{dorent2019hetero}. As illustrated in Table. \ref{tab5}, for all the tumor regions, our method achieves the best results in most of all cases ($13/15$). Compared to HeMIS, the Dice score of our method just gradually drops when modalities are missing, while the performance drop is more severe in HeMIS. Compared to WoCM, we can observe that the correlation model improved the segmentation results on average Dice Score of 5.8\%, 15.3\%, 15.6\% for whole, core and enhancing tumor, respectively. From Table. \ref{tab6}, the comparison results on BraTS 2019 dataset, we can obtain the similar observation: the correlation model can improve the segmentation results on average Dice Score of 2.4\%, 5.8\%, 19.7\% for whole, core and enhancing tumor, respectively. The correlation model makes the model more robust in the case of missing modalities, which demonstrates the effectiveness of the proposed component, and also proves our assumption that there is definitely a strong correlation in latent representation between modalities.

We also find that, missing FLAIR modality leads to a sharp decreasing on dice score for all the regions, since FLAIR is the principle modality for showing whole tumor. Missing T1 and T2 modalities would have a slight decreasing on dice score for all the regions. While missing T1c modality would have a sever decreasing on dice score for both tumor core and enhancing tumor, since T1c is the principle modality for showing tumor core and enhancing tumor regions.

The above experiment results on the two datasets indicate that the proposed network can not only achieve a promising result on a complete modalities dataset but also in the case of missing modalities. It's worth noting that, since a lack of powerful machine, our model used only 8 initial filters, we believe it can achieve better results with more filters and more powerful machine.

\begin{table*}
\centering
\caption{Evaluation of our proposed method on Brats 2018 training set, $\uparrow$ denotes the improvement compared to the previous method, bold results show the best scores for each tumor region.}
%\vspace{0.5cm}
\label{tab1}
\resizebox{15cm}{!}{%
\begin{tabular}{c|cccc|cccc|cccc}
\hline
Method & \multicolumn{4}{c}{Dice Score} & \multicolumn{4}{|c}{Hausdorff Distance} & \multicolumn{4}{|c}{Sensitivity} \\ \hline
 & Whole &  Core & Enhancing &Average  & Whole & Core & Enhancing& Average  & Whole &  Core & Enhancing &Average\\ \hline
Baseline &86.6&76.5&67.1&76.7&8.5&9.5&8.7&8.9&85.0&75.3&70.4&76.9\\
Baseline + Fusion &87.8$\uparrow$&76.5&67.8$\uparrow$&77.4$\uparrow$&7.8$\uparrow$ &9.2$\uparrow$ &8.0$\uparrow$&8.3$\uparrow$&87.3$\uparrow$&74.5&71.0$\uparrow$&77.6$\uparrow$\\
Baseline + Fusion + Reconstruction&87.9$\uparrow$&76.2&68.1$\uparrow$&77.4 &8.5&9.7 &8.3&8.8&87.4$\uparrow$&74.6$\uparrow$&71.3$\uparrow$&77.8$\uparrow$\\
Our method &\textbf{88.2$\uparrow$}&\textbf{78.6$\uparrow$}& \textbf{69.4$\uparrow$}&78.7$\uparrow$& \textbf{6.7$\uparrow$}&\textbf{7.6$\uparrow$}& \textbf{7.1}$\uparrow$&7.1$\uparrow$&\textbf{87.7}$\uparrow$&\textbf{77.3}$\uparrow$&\textbf{72.7}$\uparrow$&79.2$\uparrow$\\ 
\hline
\end{tabular}
}
\end{table*}

\begin{table*}
\centering
\caption{Evaluation of our proposed method on Brats 2019 training set, $\uparrow$ denotes the improvement compared to the previous method, bold results show the best scores for each tumor region.}
\label{tab2}
\resizebox{15cm}{!}{%
\begin{tabular}{c|cccc|cccc|cccc}
\hline
Method & \multicolumn{4}{c}{Dice Score} & \multicolumn{4}{|c}{Hausdorff Distance} & \multicolumn{4}{|c}{Sensitivity} \\ \hline
 & Whole &  Core & Enhancing &Average  & Whole & Core & Enhancing& Average  & Whole &  Core & Enhancing &Average\\ \hline
Baseline &87.8&75.7&68.8&77.4&8.1&14.0&\textbf{6.6}&9.6&83.3&68.5&63.2&71.7\\
Baseline + Fusion &88.3$\uparrow$&76.3$\uparrow$&70.0$\uparrow$&78.2$\uparrow$&8.0$\uparrow$&9.4$\uparrow$&7.3&8.2$\uparrow$&83.6$\uparrow$&70.7$\uparrow$&65$\uparrow$&73.1$\uparrow$\\
Baseline + Fusion + Reconstruction &\textbf{89.7}$\uparrow$&76.8$\uparrow$&69.1&78.5$\uparrow$&\textbf{5.2}$\uparrow$&7.6$\uparrow$&7.4&\textbf{6.7}$\uparrow$&\textbf{94.5}$\uparrow$&72.4$\uparrow$&67.6$\uparrow$&78.2$\uparrow$\\

Our method&\textbf{89.7}&\textbf{77.5}$\uparrow$&\textbf{70.6}$\uparrow$&\textbf{79.3}$\uparrow$&6.0&\textbf{7.5}$\uparrow$&7.6&7.0&93.3&\textbf{73.7}$\uparrow$&\textbf{68.7}$\uparrow$&\textbf{78.6}$\uparrow$\\

\hline
\end{tabular}
}
\end{table*}

\begin{table*}[]
\centering
\caption{Comparison of different methods on Brats 2018 validation set, fields with (-) are not mentioned in the published work, bold results show the best scores for each tumor region, and underline results refer the second best results.}
%\vspace{0.5cm}
\label{tab3}
\resizebox{15cm}{!}{%
\begin{tabular}{c|cccc|cccc|cccc}
\hline
Method & \multicolumn{4}{c}{Dice Score} & \multicolumn{4}{|c}{Hausdorff Distance}  & \multicolumn{4}{|c}{Sensitivity}\\ \hline
 & Whole & Core & Enhancing &Average& Whole & Core & Enhancing&Average & Whole & Core & Enhancing &Average\\ \hline
 
Ronneberger et al. \cite{ronneberger2015u} & 27.6 & 9.0 & 3.7 &13.4 &53.2 & 134.1 & 187.9&125.1&25.0 &6.9&2.9&11.6 \\

Tuan et al. \cite{tuan2018brain}& 81.8 & 69.9 & 68.2 &73.3& 9.4 &12.4 &7.0&9.6&77.3&64.7&70.2&70.7\\

Hu et al. \cite{hu2018brain} &88.0 & 74.0 & 69.0 &77.0& \underline{4.7} & 10.6 &\underline{6.6}&\underline{7.3}&\textbf{87.0}&\underline{77.0}&\underline{71.0}&\underline{78.3}\\

Myronenko et al. \cite{myronenko20183d}&\textbf{90.4}&\textbf{85.9}&\textbf{81.4} &\textbf{85.9} &\textbf{4.4} & \textbf{8.2} & \textbf{3.8}&\textbf{5.5} &-&-&-&-\\

Our method&\underline{87.1}&\underline{78.3}&\underline{70.8}&\underline{78.7}&6.5&\underline{9.9}&7.1&7.8&\underline{86.8}& \textbf{78.7}& \textbf{81.9}& \textbf{82.5}\\ \hline
\end{tabular}%
}
\end{table*}

\begin{table*}[]
\centering
\caption{Comparison of different methods on Brats 2019 validation set, fields with (-) are not mentioned in the published work, bold results show the best scores for each tumor region, and underline results refer the second best results.}
\label{tab4}
\resizebox{15cm}{!}{%
\begin{tabular}{c|cccc|cccc|cccc}
\hline
Method & \multicolumn{4}{c}{Dice Score} & \multicolumn{4}{|c}{Hausdorff Distance}  & \multicolumn{4}{|c}{Sensitivity}\\ \hline
 & Whole & Core & Enhancing &Average& Whole & Core & Enhancing&Average & Whole & Core & Enhancing &Average\\ \hline
 
Agravat et al.\cite{agravat2019brain} &70.0&63.0&60.0&64.3&14.3&17.1&11.6&14.3&63.0&61.0&59.0&61.0\\
 
Boutry et al. \cite{boutry2019using} &68.4&\textbf{87.8}&\textbf{74.7}&77.0&10.2&\underline{10.9}&14.8&12.0&-&-&-&-\\

Kim et al. \cite{kim2019two}&\textbf{87.6}&76.4&67.2&77.1&14.1&11.6&8.8&11.5&\textbf{88.7}&\textbf{76.5}&\textbf{76.3}&\textbf{80.5}\\

Amian et al. \cite{amian2019multi}&86.0&\underline{77.0}&71.0&\textbf{78.0}&\underline{8.4}&11.5&\underline{6.9}&\underline{8.9}&85.0&\underline{76.0}&69.0&\underline{76.7}\\

Our method &\underline{87.1}&71.8&\underline{72.7}&\underline{77.2}&\textbf{6.7}&\textbf{9.3}&\textbf{6.3}&\textbf{7.4}&\underline{85.6}&67.7&\underline{75.2}&76.2\\ \hline
\end{tabular}%
}
\end{table*}

\begin{table*}[]
\centering
\caption{Robust comparison of different methods (Dice \%) for different combinations of available modalities on BraTS 2018 dataset, $\circ$ denotes the missing modality and $\bullet$ denotes the present modality, $\uparrow$ denotes the improvement of CM, WoCM denotes our method without CM, Our denotes our method with CM, bold results denotes the best score.}
%\vspace{0.5cm}
\label{tab5}
\resizebox{15cm}{!}{%
\begin{tabular}{cccc|ccccccccc}
\hline
\multicolumn{4}{c|}{Modalities} & \multicolumn{3}{c|}{Whole Tumor} & \multicolumn{3}{c|}{Tumor Core} & \multicolumn{3}{c}{Enhancing Tumor} \\ \hline
F & T1 & T1c & T2 & WoCM & Our & \multicolumn{1}{c|}{HeMIS} & WoCM & Our & \multicolumn{1}{c|}{HeMIS} & WoCM & Our & HeMIS \\ \hline
$\circ$ & $\circ$ & $\circ$ & $\bullet$ & 31.4 & 33.0 $\uparrow$ & \multicolumn{1}{c|}{\textbf{38.6}} & 14.9 &15.9 $\uparrow$ & \multicolumn{1}{c|}{\textbf{19.5}} & 6.2 & \textbf{7.2 $\uparrow$} & 0.0 \\
$\circ$ & $\circ$ & $\bullet$ & $\circ$ & 29.7 & \textbf{33.6 $\uparrow$} & \multicolumn{1}{c|}{2.6} & 49.3 & \textbf{56.1 $\uparrow$} & \multicolumn{1}{c|}{6.5} & 50.0 & \textbf{53.5 $\uparrow$} & 11.1 \\
$\circ$ & $\bullet$ & $\circ$ & $\circ$ & 3.3 & \textbf{5.6 $\uparrow$} & \multicolumn{1}{c|}{0.0} & 4.3 & \textbf{6.3 $\uparrow$} & \multicolumn{1}{c|}{0.0} & 4.5 & \textbf{5.3 $\uparrow$} & 0.0 \\
$\bullet$ & $\circ$ & $\circ$ & $\circ$ & 71.4 & \textbf{73.7 $\uparrow$} & \multicolumn{1}{c|}{55.2} & 46.2 & \textbf{48.6 $\uparrow$} & \multicolumn{1}{c|}{16.2} & 5.0 & \textbf{25.8 $\uparrow$} & 6.6 \\
$\circ$ & $\circ$ & $\bullet$ & $\bullet$ & 45.1 & \textbf{48.3 $\uparrow$} & \multicolumn{1}{c|}{48.2} & 48.1 & \textbf{50.4 $\uparrow$} & \multicolumn{1}{c|}{45.8} & 52.0 & 52.4 $\uparrow$ & \textbf{55.8} \\
$\circ$ & $\bullet$ & $\bullet$ & $\circ$ & 11.4 & \textbf{29.2 $\uparrow$} & \multicolumn{1}{c|}{15.4} & 22.6 & \textbf{55.0 $\uparrow$} & \multicolumn{1}{c|}{30.4} & 24.8 & \textbf{54.8 $\uparrow$} & 42.6 \\
$\bullet$ & $\bullet$ & $\circ$ & $\circ$ & 75.9 & \textbf{80.4 $\uparrow$} & \multicolumn{1}{c|}{71.1} & 47.4 & \textbf{51.5 $\uparrow$} & \multicolumn{1}{c|}{11.9} & 7.7 & \textbf{10.2 $\uparrow$} & 1.2 \\
$\circ$ & $\bullet$ & $\circ$ & $\bullet$ & 31.6 & 35.5 $\uparrow$ & \multicolumn{1}{c|}{\textbf{47.3}} & 12.9 & 14.3 $\uparrow$ & \multicolumn{1}{c|}{\textbf{17.2}} & 2.5 & \textbf{6.1 $\uparrow$} & 0.6 \\
$\bullet$ & $\circ$ & $\circ$ & $\bullet$ & 80.4 & \textbf{81.3 $\uparrow$} & \multicolumn{1}{c|}{74.8} & 20.7 & \textbf{25.2 $\uparrow$} & \multicolumn{1}{c|}{17.7} & 9.3 & \textbf{10.0 $\uparrow$} & 0.8 \\
$\bullet$ & $\circ$ & $\bullet$ & $\circ$ & 80.3 & \textbf{81.5 $\uparrow$} & \multicolumn{1}{c|}{68.4} & 65.7 & \textbf{73.4 $\uparrow$} & \multicolumn{1}{c|}{41.4} & 62.7 & \textbf{67.5 $\uparrow$} & 53.8 \\
$\bullet$ & $\bullet$ & $\bullet$ & $\circ$ & 81.1 & \textbf{82.7 $\uparrow$} & \multicolumn{1}{c|}{70.2} & 71.7 & \textbf{75.8 $\uparrow$} & \multicolumn{1}{c|}{48.8} & 65.7 & \textbf{68.4 $\uparrow$} & 60.9 \\
$\bullet$ & $\bullet$ & $\circ$ & $\bullet$ & 83.5 & \textbf{85.4 $\uparrow$} & \multicolumn{1}{c|}{75.2} & 41.3 & \textbf{44.4 $\uparrow$} & \multicolumn{1}{c|}{18.7} & 11.1 & \textbf{12.9 $\uparrow$} & 1.0 \\
$\bullet$ & $\circ$ & $\bullet$ & $\bullet$ & 87.5 & \textbf{87.9 $\uparrow$} & \multicolumn{1}{c|}{75.6} & 74.2 & \textbf{77.5 $\uparrow$} & \multicolumn{1}{c|}{54.9} & 65.4 & \textbf{67.2 $\uparrow$} & 60.5 \\
$\circ$ & $\bullet$ & $\bullet$ & $\bullet$ & 46.9 & \textbf{50.1 $\uparrow$} & \multicolumn{1}{c|}{44.2} & 51.2 & \textbf{52.1 $\uparrow$} & \multicolumn{1}{c|}{46.6} & 54.3 & 54.8 $\uparrow$ & \textbf{55.1} \\
$\bullet$ & $\bullet$ & $\bullet$ & $\bullet$ & 87.9 & \textbf{88.2 $\uparrow$} & \multicolumn{1}{c|}{73.8} & 76.2 & \textbf{78.6 $\uparrow$} & \multicolumn{1}{c|}{55.3} & 68.1 & \textbf{69.4 $\uparrow$} & 61.1 \\ \hline
\multicolumn{4}{c|}{Wins / 15} & 0 & 13 & \multicolumn{1}{c|}{2} & 0 & 13 & \multicolumn{1}{c|}{2} & 0 & 13 & 2 \\ \hline
\end{tabular}%
}
\end{table*}

\begin{table*}[]
\centering
\caption{Robust comparison of different methods (Dice \%) for different combinations of available modalities on BraTS 2019 dataset, $\circ$ denotes the missing modality and $\bullet$ denotes the present modality, $\uparrow$ denotes the improvement of CM, WoCM denotes our method without CM, Our denotes our method with CM, bold results denotes the best score.}
%\vspace{0.5cm}
\label{tab6}
\resizebox{12cm}{!}{%
\begin{tabular}{cccc|cc|cc|cc}
\hline
\multicolumn{4}{c|}{Modalities} & \multicolumn{2}{c|}{Whole Tumor} & \multicolumn{2}{c|}{Tumor Core} & \multicolumn{2}{c}{Enhancing Tumor} \\ \hline
F & T1 & T1c & T2 & WoCM & Our & WoCM & Our & WoCM & Our \\ \hline
$\circ$ & $\circ$ & $\circ$ & $\bullet$ &62.9&\textbf{70.3}$\uparrow$&19.6&\textbf{32.6}$\uparrow$&\textbf{3.8}&\textbf{3.8}\\

$\circ$ & $\circ$ & $\bullet$ & $\circ$ &\textbf{20.0}&\textbf{20.0}&39.8&\textbf{41.7}$\uparrow$&36.2&\textbf{46.8}$\uparrow$\\

$\circ$ & $\bullet$ & $\circ$ & $\circ$ &8.0&\textbf{10.0}$\uparrow$&0.8&\textbf{2.2}$\uparrow$&0.9&\textbf{1.0}$\uparrow$\\

$\bullet$ & $\circ$ & $\circ$ & $\circ$ &71.9&\textbf{74.3}$\uparrow$&48.1&\textbf{52.0}$\uparrow$&13.8&\textbf{17.0}$\uparrow$\\

$\circ$ & $\circ$ & $\bullet$ & $\bullet$ &72.3&\textbf{73.5}$\uparrow$&63.7&\textbf{66.8}$\uparrow$&48.4&\textbf{60.4}$\uparrow$ \\

$\circ$ & $\bullet$ & $\bullet$ & $\circ$&18.4&\textbf{20.7}$\uparrow$&36.7&\textbf{39.3}$\uparrow$&39.7&\textbf{42.3}$\uparrow$\\

$\bullet$ & $\bullet$ & $\circ$ & $\circ$ &80.6&\textbf{81.6}$\uparrow$&52.7&\textbf{55.6}$\uparrow$&1.8&\textbf{3.0}$\uparrow$\\

$\circ$ & $\bullet$ & $\circ$ & $\bullet$& 70.0&\textbf{71.6}$\uparrow$&38.3&\textbf{42.1}$\uparrow$&1.0&\textbf{1.2}$\uparrow$\\

$\bullet$ & $\circ$ & $\circ$& $\bullet$ &80.6&\textbf{83.9}$\uparrow$&41.1&\textbf{41.8}$\uparrow$&6.9&\textbf{11.3}$\uparrow$\\

$\bullet$ & $\circ$ & $\bullet$ & $\circ$&\textbf{81.7}&\textbf{81.7}&72.6&\textbf{74.7}$\uparrow$&48.7&\textbf{63.5}$\uparrow$\\

$\bullet$ & $\bullet$ & $\bullet$&$\circ$&83.0&\textbf{83.9}$\uparrow$&75.2&\textbf{75.7}$\uparrow$&55.5&\textbf{67.1}$\uparrow$\\

$\bullet$ & $\bullet$ & $\circ$ & $\bullet$ &87.7&\textbf{88.2}$\uparrow$&50.3&\textbf{52.1}$\uparrow$&1.7&\textbf{2.9}$\uparrow$\\

$\bullet$ &$\circ$ &$\bullet$  &$\bullet$&88.7&\textbf{88.9}$\uparrow$&73.6&\textbf{76.6}$\uparrow$&52.2&\textbf{64.4}$\uparrow$\\

$\circ$&$\bullet$ & $\bullet$&$\bullet$&73.5&\textbf{74.2}$\uparrow$&64.6&\textbf{66.6}$\uparrow$&53.1&\textbf{64.0}$\uparrow$\\

$\bullet$ & $\bullet$ & $\bullet$ & $\bullet$ & \textbf{89.7}&\textbf{89.7}&76.8&\textbf{77.5}$\uparrow$ &69.1&\textbf{70.6}$\uparrow$ \\
\hline
\multicolumn{4}{c|}{Wins / 15} & 3 & 15& 0 & 15 & 1 & 15\\ \hline
\end{tabular}%
}
\end{table*}

\subsection{Qualitative analysis}
\label{4.2}
\begin{figure*}[htb]
\centering
\includegraphics[width=15cm]{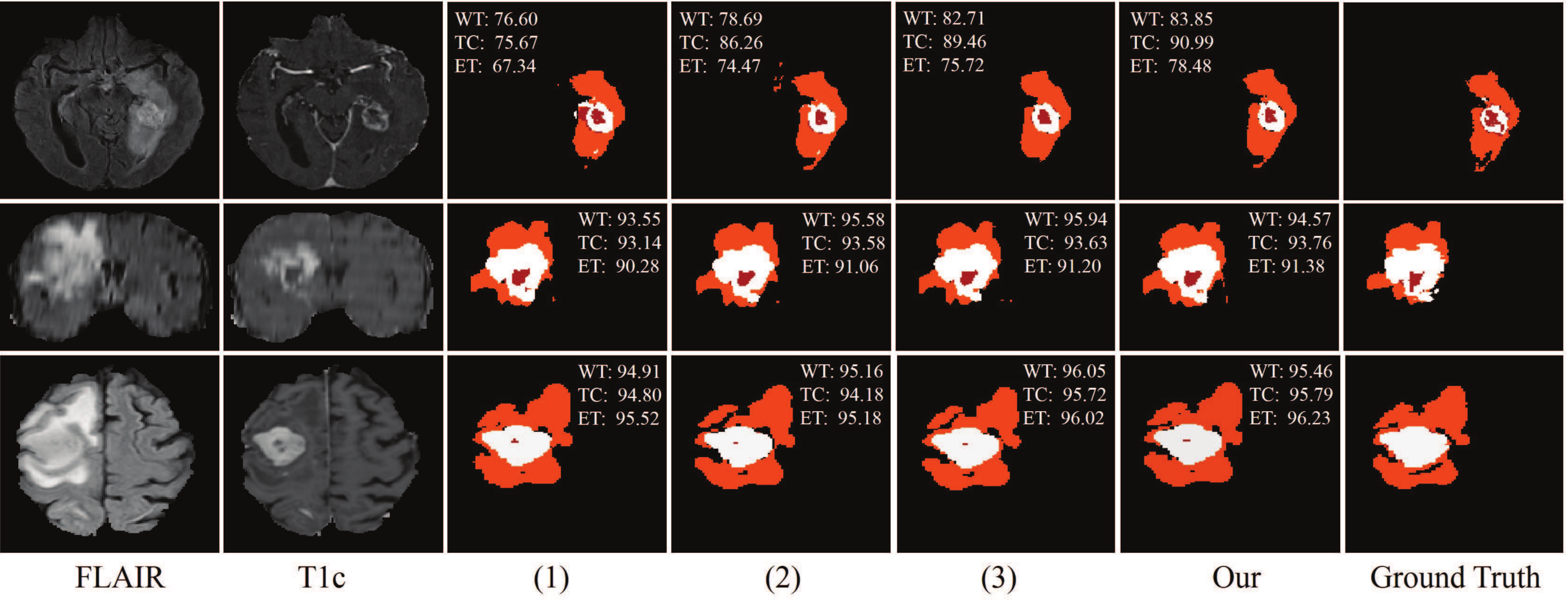}
\caption{Examples of the segmentation results on full modalities on BraTS 2018 dataset. (1) denotes the baseline, (2) denotes the baseline with fusion block, (3) denotes our method without CM. Red: necrotic and non-enhancing tumor core; Orange: edema; White: enhancing tumor.}
\label{fig7}
\end{figure*}

\begin{figure*}[htb]
\centering
\includegraphics[width=15cm]{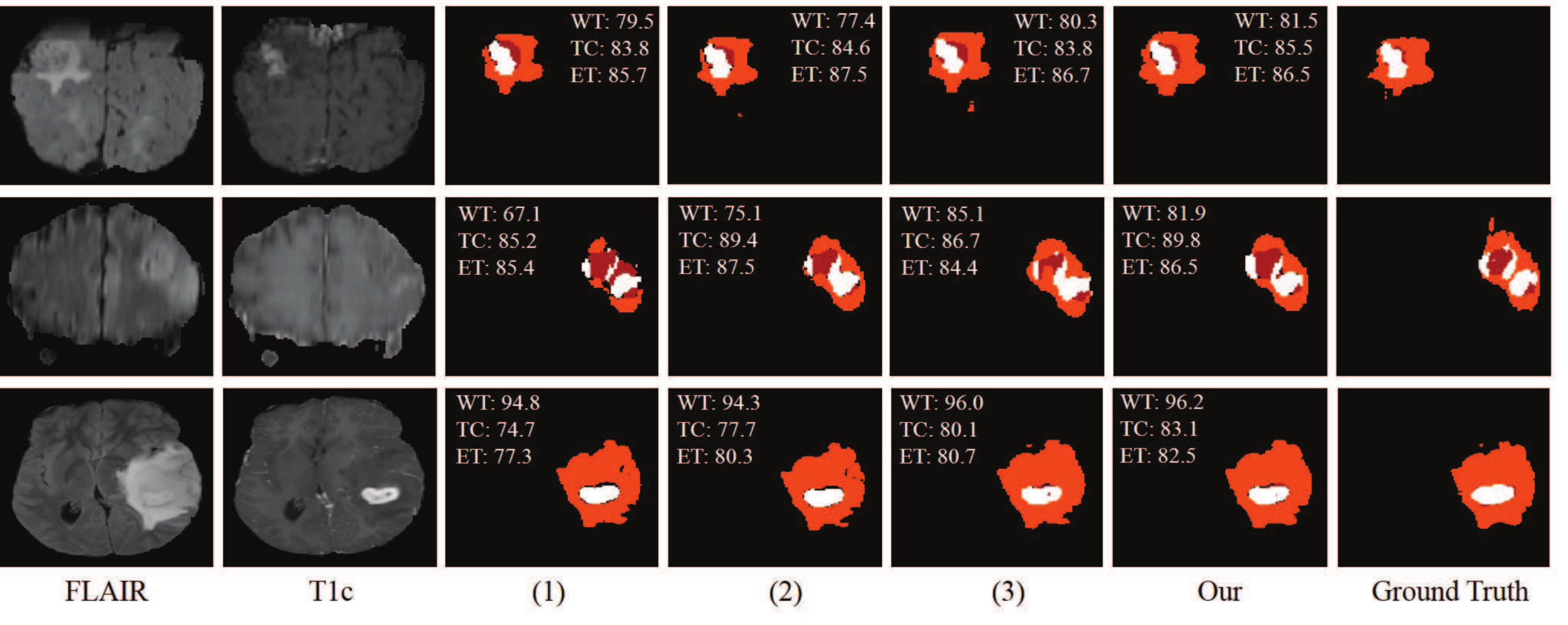}
\caption{Examples of the segmentation results on full modalities on BraTS 2019 dataset. (1) denotes the baseline, (2) denotes the baseline with fusion block, (3) denotes our method without CM. Red: necrotic and non-enhancing tumor core; Orange: edema; White: enhancing tumor.}
\label{fig8}
\end{figure*}

\begin{figure*}[htb]
\centering
\includegraphics[width=15cm]{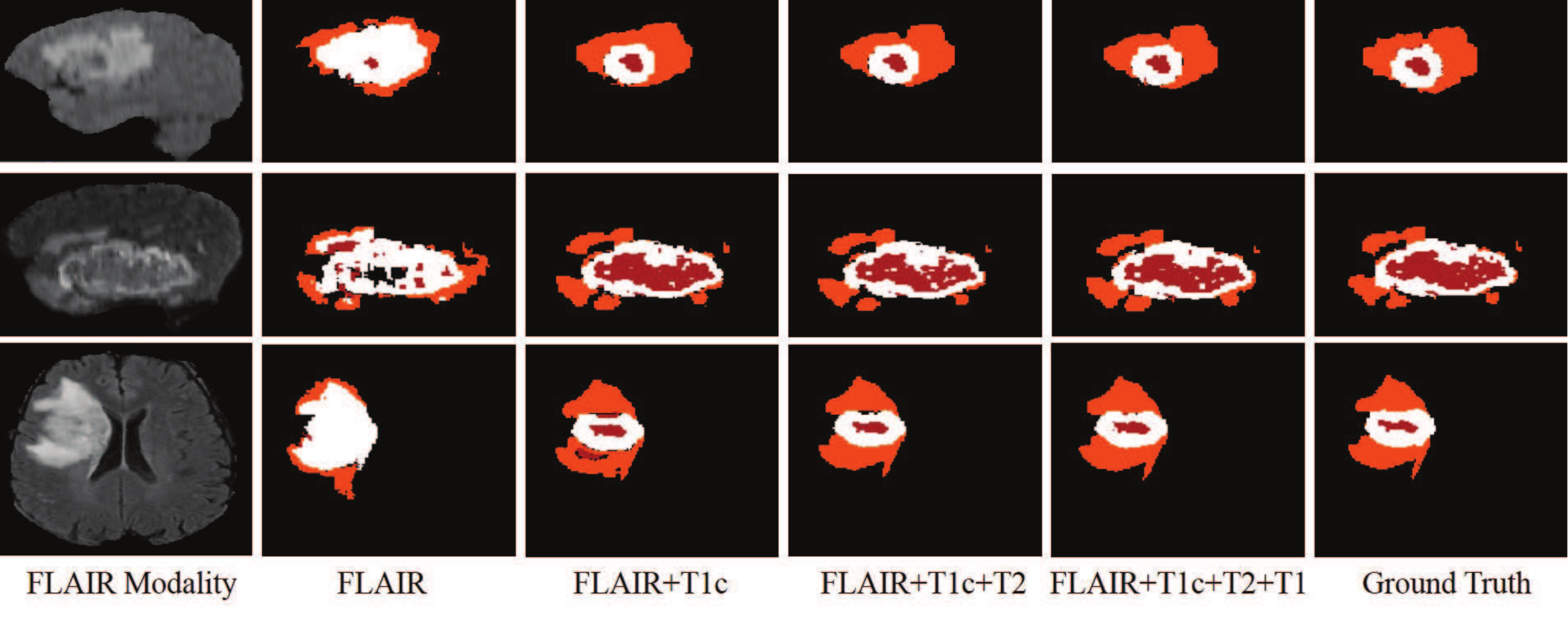}
\caption{Examples of the segmentation results on missing modalities. Red: necrotic and non-enhancing tumor core; Orange: edema; White: enhancing tumor.}
\label{fig9}
\end{figure*}

In order to evaluate the robustness of our model, we randomly select several examples from BraTS 2018 and BraTS 2019 dataset, and visualize the segmentation results of different methods on full modalities in Fig. \ref{fig7} and Fig. \ref{fig8}. The segmentation results on missing modalities are presented in Fig. \ref{fig9}.

From Fig. \ref{fig7} and Fig. \ref{fig8}, We can observe that the segmentation results are gradually improved when the proposed strategies are integrated, these comparisons indicate that the effectiveness of the proposed strategies. In addition, with all the proposed strategies, our proposed method can achieve almost the same results with ground truth.

As shown in Fig. \ref{fig9}, we can observe that with the increasing number of missing modalities, the segmentation results produced by our robust model just slightly degrade, rather than a sudden sharp degrading. In addition, only using FLAIR modality, the proposed method can generate a good segmentation of whole tumor. And with FLAIR and T1c modalities, it can yield a competitive segmentation result compared to the ground truth. We can also find that, T1 and T2 modalities can help to refine the boundary area of the tumor regions to achieve the best segmentation.

\begin{figure}[htb]
\centering
\includegraphics[width=8cm]{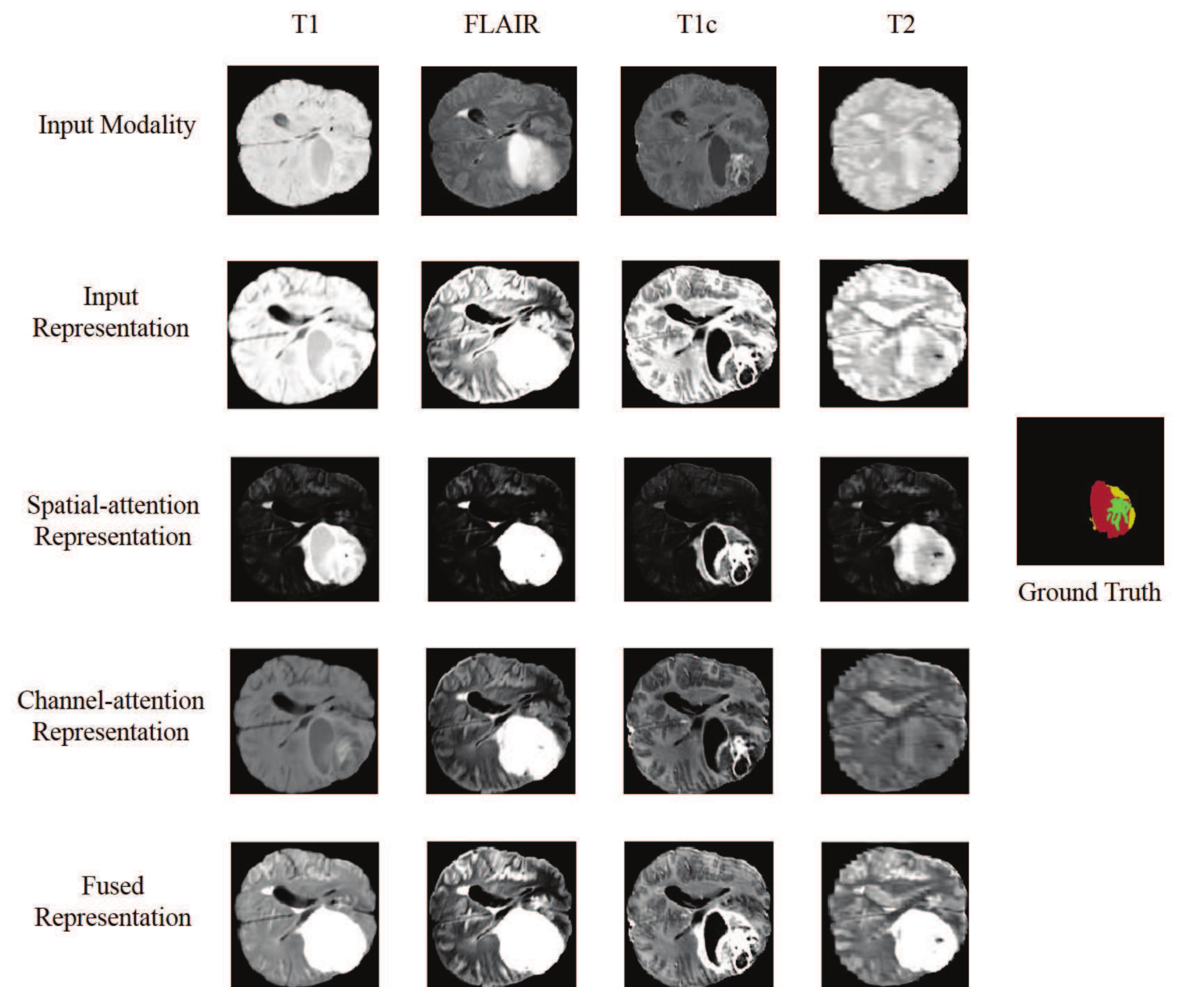}
\caption{Visualization of feature maps of the attention based fusion block. Red: necrotic and non-enhancing tumor core; Yellow: edema; Green: enhancing  tumor.}
\label{fig10}
\end{figure}

\subsection{Visualization of the attention based fusion block}
\label{4.3}
To show the performance of our proposed attention based fusion block, we select an example on BraTS 2018 dataset to see the feature maps of the first layer in our network. From Fig.~\ref{fig10}, we can observe the characteristics of the target tumors in the four independent feature representations ($F_1$, $F_2$, $F_3$, $F_4$) are not obvious. However, the spatial attention module can emphasize the position of the tumor regions (see Fig.~\ref{fig10}, third row), and the channel attention module can stand out the strength of each modalities (see Fig.~\ref{fig10}, fourth row). For example, the FLAIR focuses on the whole tumor region and T1c stands out the tumor core regions (red and green regions). Finally, by combining the spatial-attention and channel-attention feature representations, the interested tumor regions can be clearly identified in the fused feature representations.

\section{Discussion and Conclusion}
\label{sec5}
In this paper, we have presented a novel brain tumor segmentation network to deal with the absence of imaging modalities. To the best of our knowledge, this is the first segmentation method which is capable of describing the latent multi-source correlation representation between modalities and allows to help segmentation on missing modalities. Since most current segmentation networks are single-encoder based and thus can't take advantage of the correlated information available between different modalities. To this end, we designed a multi-encoder based segmentation network composed of four parts: encoder, correlation model, fusion block and decoder. The encoders are designed to obtain the individual feature representations from the input images, to learn the contributions of the obtained feature representations for the segmentation, we propose a fusion block based on attention mechanism, which allows to selectively emphasize feature representations along channel attention and spatial attention. In addition, the proposed correlation model is used to discover the latent correlations between the feature representations of the four modalities, which making the segmentation robust when the modalities are gradually missing. We carried out extensive experiments to evaluate our proposed method. The quantitative and qualitative analysis on BraTS 2018 and BraTS 2019 dataset demonstrate the effectiveness of our proposed method.

To analyze the impact of the proposed components of our network, several ablation experiments are implemented with regard to the fusion block, reconstruction decoder, and correlation model. The comparison results demonstrate the proposed strategies can aide the network gradually refine the segmentation results until to achieve the best results. We also compared our method with the state-of-the-art approaches on full and missing modalities on BraTS 2018 and BraTS 2019 dataset. Although we didn't surpass the best method on full modalities, while our results are still competitive for the segmentation. Since this work is presented to cope with the segmentation on missing modalities, and both the quantitative and qualitative results have demonstrated that our method can achieve a better result compared to the state-of-the-art method when the modalities are gradually missing.

In the future, we would like to investigate more complex
model to describe the multi-source correlation representation and adapt it to missing data issue.

\section*{Acknowledgments}
This work was co-financed by the European Union with the European regional development fund (ERDF, 18P03390/18E01750/18P02733) and by the Haute-Normandie Regional Council via the M2SINUM project. This work was partly supported by the China Scholarship Council (CSC).

\ifCLASSOPTIONcaptionsoff
  \newpage
\fi

%\begin{thebibliography}
\bibliography{IEEEabrv,refer}
%\end{thebibliography}

\begin{IEEEbiography}[{\includegraphics[width=1in,height=1.25in,clip,keepaspectratio]{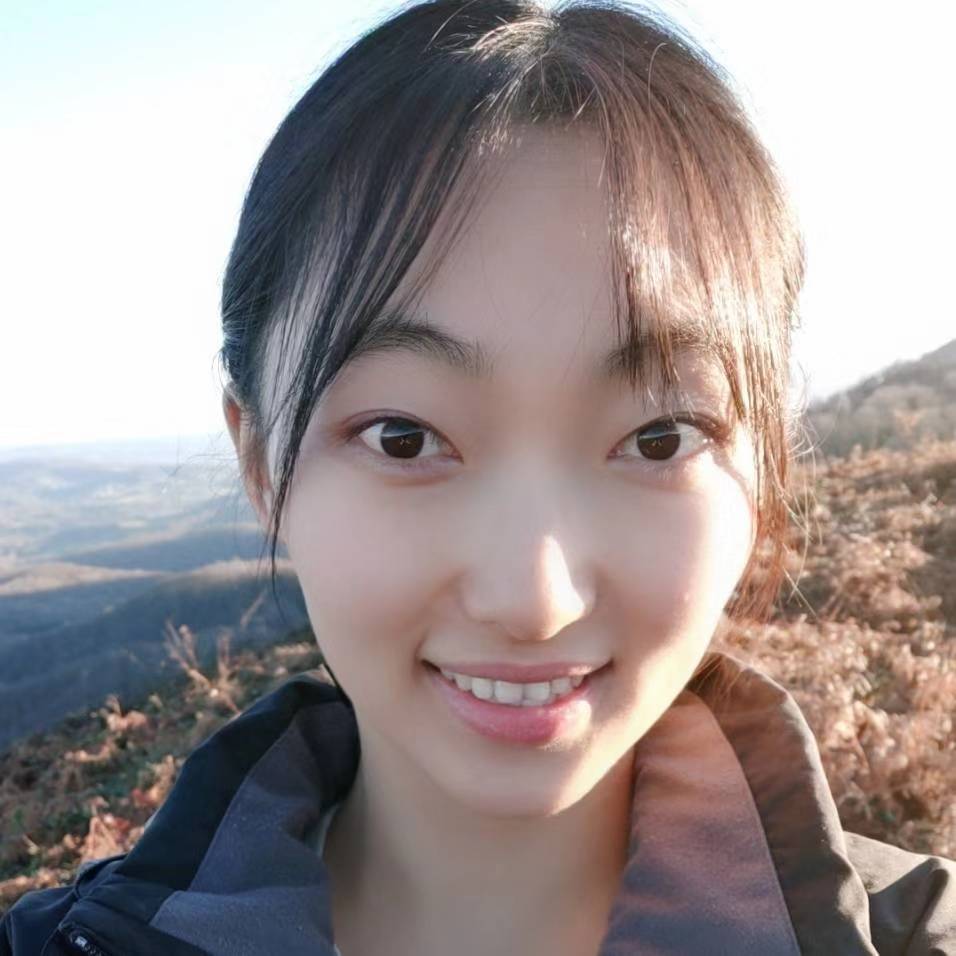}}]{Tongxue Zhou}
received the B.S. degree in Biomedical Engineering from Jilin University, Changchun, China in 2015. She is currently pursuing the Ph.D. degree in Computer Science from National Institute of Applied Sciences of Rouen (INSA Rouen Normandie), Rouen, France. Her current research interests include medical image analysis, data fusion and deep learning.
\end{IEEEbiography}

\begin{IEEEbiography}[{\includegraphics[width=1in,height=1.25in,clip,keepaspectratio]{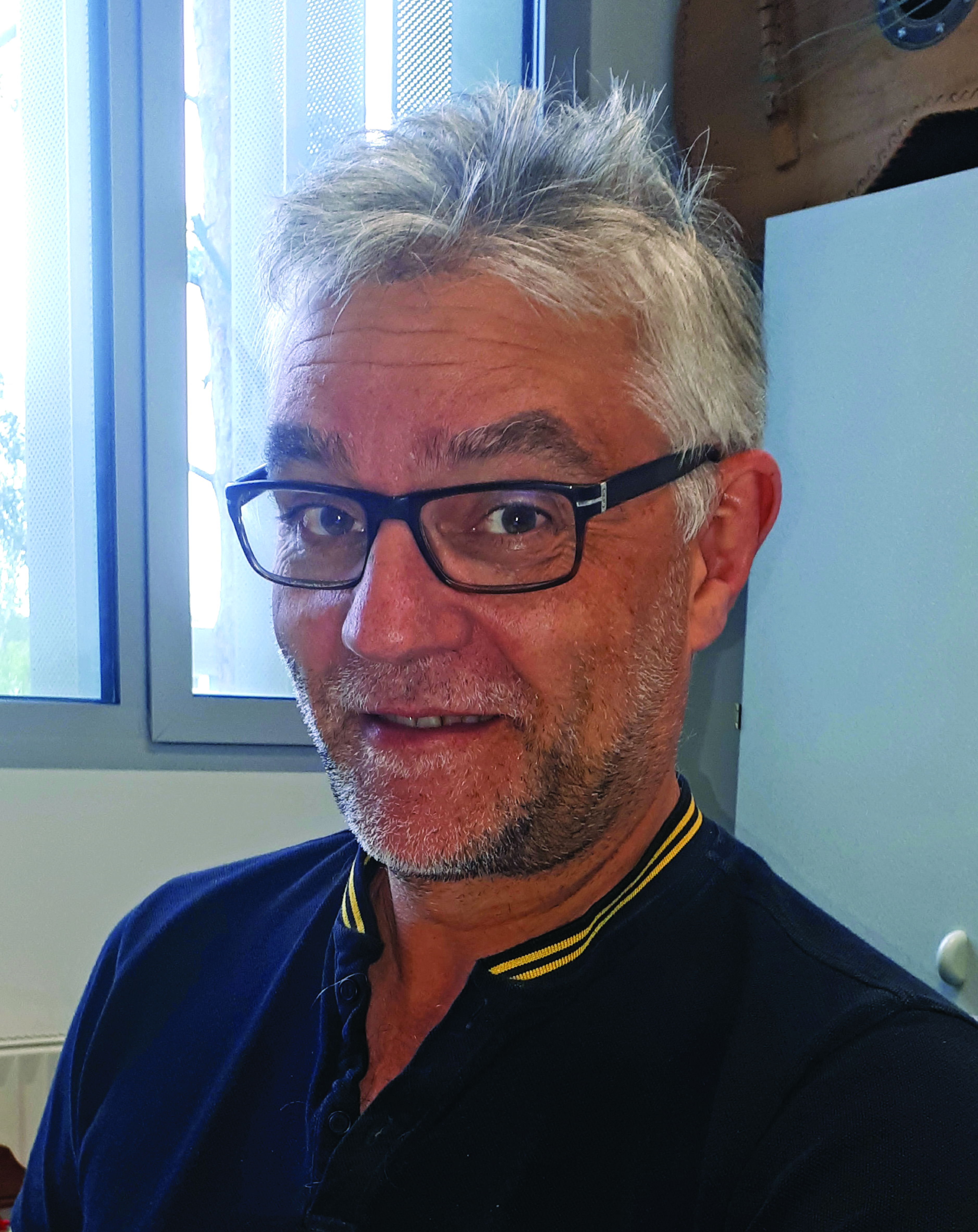}}]{Stéphane Canu}
received the Ph.D. degree in system command from the Compiègne University of Technology in 1986. He received the French Habilitation degree from Paris 6 University. He is currently a Professor of the LITIS research laboratory and of the information technology department, at the National Institute of Applied Sciences of Rouen (INSA Rouen Normandie). His research interests includes deep learning, kernels machines, regularization, machine learning applied to signal processing, pattern classification and optimization for machine learning.
\end{IEEEbiography}

\begin{IEEEbiography}[{\includegraphics[width=1in,height=1.25in,clip,keepaspectratio]{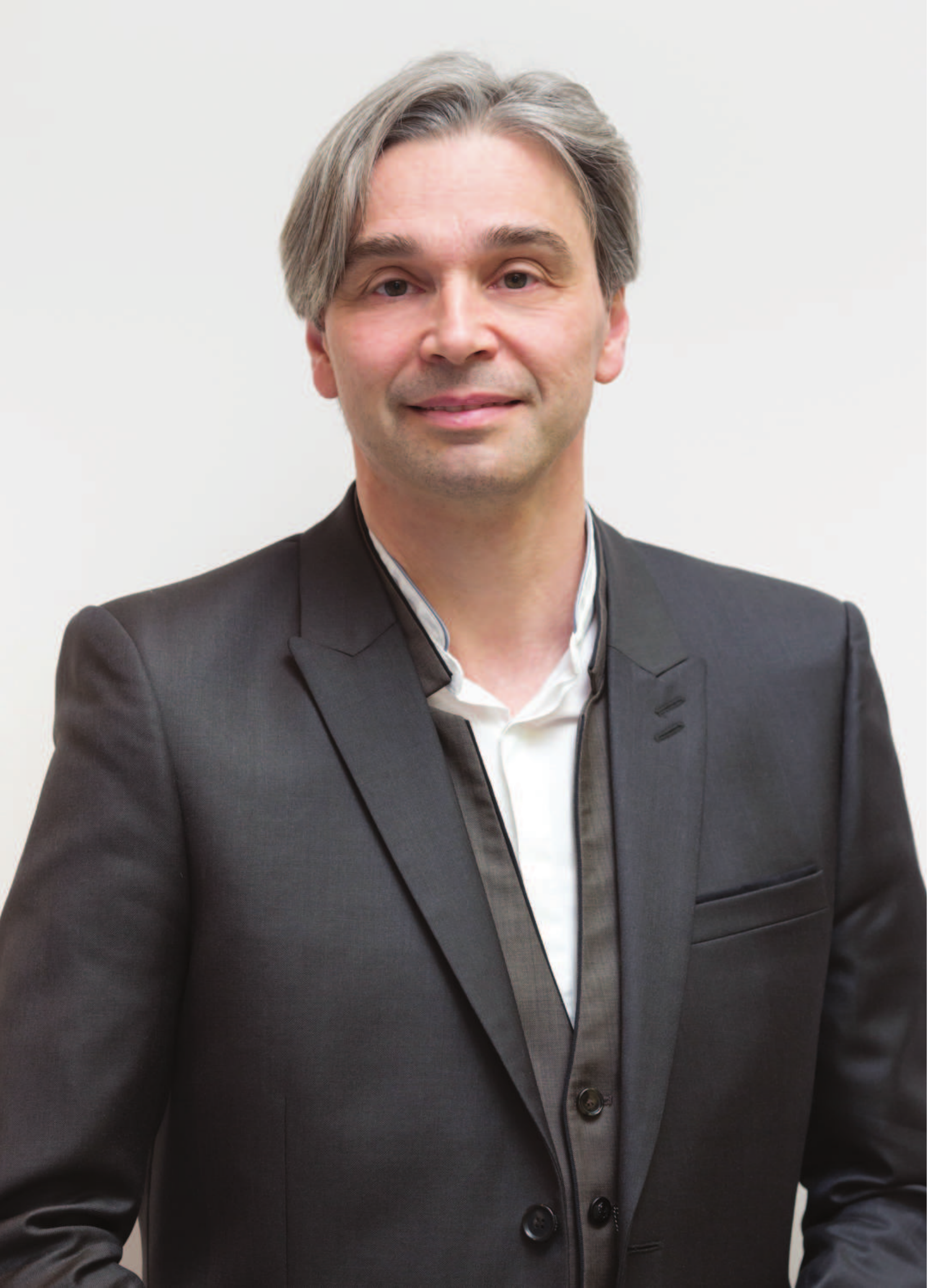}}]{Pierre Vera}
received the M.D. degree in 1993 from Université Paris VI and the PhD degree from the same institution in 1999. He is currently a University Professor and Hospital Physician with the Faculty of Medicine, University of Rouen, France. He is also the General Director of Henri Becquerel Cancer Center, and the head of the Department of Nuclear Medicine. His research interests include radiation oncology, nuclear medicine, biophysics, and medical imaging.
\end{IEEEbiography}

\begin{IEEEbiography}[{\includegraphics[width=1in,height=1.25in,clip,keepaspectratio]{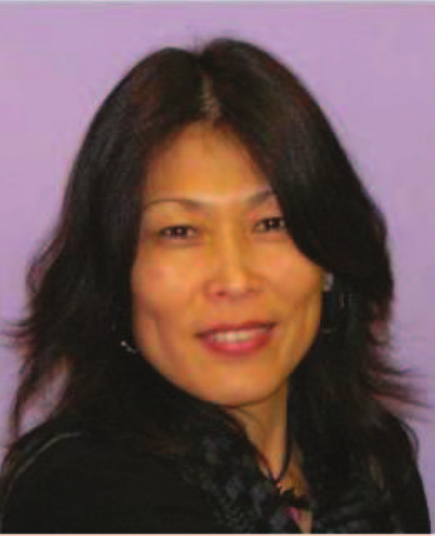}}]{Su Ruan}
received the M.S. and the Ph.D. degrees in Image Processing from the University of Rennes, France, in 1989 and 1993, respectively. She was a Full Professor in the University of Champagne-Ardenne, France, from 2003 to 2010. She is currently a Full Professor with the Department
of Medicine, and the Leader of the QuantIF Team, LITIS Research Laboratory, University of Rouen, France. Her research interests include pattern recognition, machine learning, information fusion, and medical imaging.
\end{IEEEbiography}

\end{document}